\documentclass[fleqn,usenatbib,useAMS]{mnras}
\usepackage{graphicx}
\usepackage{amsmath}
\usepackage{amssymb}
\usepackage{bm}
\usepackage{color}

\title[PDF of a Molecular Clouds ensemble]{Density distribution function of a self-gravitating isothermal compressible turbulent fluid in the context of Molecular Clouds ensembles}

\author[Donkov, Stefanov]
{\parbox{\textwidth}{Sava Donkov$^{1}$, Ivan Stefanov$^{1}$}\vspace{0.4cm} \\
  $^1$Depatment of Applied Physics, Faculty of Applied Mathematics, Technical University, 8 Kliment Ohridski Blvd., 1000 Sofia, Bulgaria \\}

\begin{document} 
\label{firstpage}

\date{Accepted XXX. Received YYY; in original form ZZZ}
\pagerange{\pageref{firstpage}--\pageref{lastpage}} \pubyear{2017}
\maketitle

\begin{abstract}
We have set ourselves the task of obtaining the probability distribution function of the mass density of a self-gravitating isothermal compressible turbulent fluid from its physics. We have done this in the context of a new notion: the molecular clouds ensemble. We have applied a new approach that takes into account the fractal nature of the fluid. Using the medium equations, under the assumption of steady state, we show that the total energy per unit mass is an invariant with respect to the fractal scales. As a next step we obtain a nonlinear integral equation for the dimensionless scale $Q$  which is the third root of the integral of the probability distribution function. It is solved approximately up to the leading-order term in the series expansion. We obtain two solutions. They are power-law distributions with different slopes: the first one is -1.5 at low densities, corresponding to a equilibrium between all energies at a given scale, and the second one is -2 at high densities, corresponding to a free fall at small scales.
\end{abstract} 


\section{Introduction}   \label{Sec-Intr}

Molecular clouds (MCs) are the birthplaces of stars. Their structure and evolution determine the initial stellar mass function (IMF) \citep{Elme_00, Offner_ea_14}. On its side the IMF determines the structure and the evolution of the Galaxy \citep{Elme_00, Offner_ea_14}.

The structure and the evolution of MCs is governed by their physics the main ingredients of which are: supersonic turbulence, self-gravity, gas thermodynamics and magnetic fields \citep{VS_10, KG_14}.
The stars born by the molecular cloud act in return on the parent cloud and change its structure and evolution. In order to simplify the picture we will neglect the magnetic field and the back-reaction of the stars on the cloud. In other words in the current study the MCs are modeled as a self-gravitating, isothermal, compressible turbulent fluid. Our main task is to obtain the probability distribution function (PDF) of the mass density starting from first principles, namely from the physics mentioned above.

Why is this important? As we have already seen the structure and the evolution of the MCs determine the IMF \citep{Elme_00, Offner_ea_14}.
The evolution of the cloud is determined by the cloud's structure, through the initial conditions, and by the equations of the medium \citep{VS_10, HF_12}. The structure of the molecular cloud is encoded in its PDF \citep{HF_12}. The latter should be understood in statistical sense, not literary -- in morphological sense. The stochastic character of the cloud is revealed in the fact that it has fractal structure \citep{Elme_97}. The fractalness implies a hierarchical structure which is most suitably described in terms of abstract scales.

So, we are confronted with the task to obtain the PDF of the mass density from the equations of the medium assuming that the medium, i.e. the cloud, is a hierarchical (fractal) structure.

As it is shown in  \citet{deVSC_98}, the fractal structure and the scaling laws observed in self-gravitating gases are a consequence of the physical equations of the medium. Let us note again that we neglect the magnetic field and the back-reaction of the newborn stars and assume that the medium is isothermal \citep{Ferriere_01}. As stated bellow in Section \ref{subsec-model physics}, we make also the following assumptions: with respect to the macroscopic motion of the fluid elements (governed by the equations of the medium) the medium is locally homogeneous and isotropic. The system as a whole (the MC) is in steady state. The extent to which these assumptions are relevant for real physical objects is commented on in Section \ref{subsec-model physics} and in the discussion (Section \ref{Sec-discussion}).

The paper has the following structure.  Section \ref{Sec-Model} presents the setup of the model. The general physical picture is given in \ref{subsec-model physics}, while the geometry of the abstract object that we work with is in \ref{subsec-model geometry}. Section \ref{Sec-Equ_p(s)} is devoted to the derivation of the equation  for the PDF of the mass density. This is the core of the outline. The derivation is split into several steps: in \ref{subsec-equ_p(s)_general} as a consequence of the general construction and the compressible Euler equations it is shown that the total energy per unit mass of a fluid element is invariant with respect to the fractal scales; in \ref{subsec-equ_p(s)_terms} the explicit form of the different terms of the invariant is discussed; in \ref{subsec-equ_p(s)_accretion term} with the help of the continuity equation the accretion velocity is obtained in explicit form; finally, in \ref{subsec-equ_p(s)_diff equ}, we obtain an equation for the PDF of mass density. It is studied in Section~\ref{Sec-analysis and sol} where we give also an approximate solution in the form of exponential functions which represents the leading-order term in the series expansion. In Section \ref{Sec-discussion} we comment on the assumptions of the model and on the results, compare them to similar models and solutions from the past and to results from simulations and observations. At the end, we propose directions for future research. The conclusion is in Section \ref{Sec-Concl}.

\section{Setup of the model}
\label{Sec-Model}

\subsection{General physical picture}
\label{subsec-model physics}

Let us consider a self-gravitating, turbulent fluid with supersonic turbulent velocity fluctuations. The speed of sound depends on the temperature of the medium. The medium is assumed to be an isothermal ideal gas, locally (for volumes which are small with respect to the entire system) described by the Clapeyron-Mendeleev equation of state:

\begin{equation}
\label{Cl-M_cs}
p_{\rm th}=\rho \frac{RT}{\mu} ~~ \Rightarrow ~~ c_{\rm s}^2=\frac{dp_{\rm th}}{d\rho}=\frac{RT}{\mu}~,
\end{equation}
where $p_{\rm th}$ is the local gas pressure produced by the thermal motion of the gas molecules; $\rho$ is the average local gas density; $T$ is the local gas temperature, which we assume to have the same value in the whole volume of the gas; $\mu$ is the molar mass; $R$ is the universal gas constant, and $c_{\rm s}$ is the speed of sound. By saying that the Clapeyron-Mendeleev equation is valid only locally we mean that the pressure and the density of the gas take different values from one region to the other due to self-gravity \citep{deVSC_98}. Our considerations are only for regions which are small enough to be treated as homogeneous with respect to  $p_{\rm th}$ and $\rho$.

We assume that our physical system, the medium, is at both microscopic and macroscopic equilibrium. The former means that the medium is isothermal. The latter means that the macroscopic motion of the fluid elements which is driven by turbulence and gravity is quasi-static. In other words, from statistical point of view the gas is in steady state at both microscopic and macroscopic level. Macroscopic steady state is equivalent to say that the density distribution does not evolve with time.

We assume that the turbulence is saturated, locally homogeneous and isotropic. Besides, the scales are in the inertial region which means that energy is introduced in the system at scales much greater than its size. According to some general theoretical considerations, which are supported also by numerical experiments \citep{deVSC_96a, deVSC_96b, deVSC_98}, such a physical system can be treated as a fractal which yields scaling laws for the mass and the turbulent velocity fluctuations.

Given this setup we expect that the gas density averaged over long enough time intervals would vary from one point to the other due to supersonic turbulence and gravity. Compressible turbulence alone can produce large enough density fluctuations so that a tangible density distribution is observed in the time-averaged picture. In the case of an isothermal system the distribution of the logarithm of the density caused by compressible turbulence is Gaussian, i.e. the so called log-normal distribution \citep{VS_94, NP_99, Kritsuk_ea_07, Federrath_ea_10}. The joint action of supersonic turbulence and gravity, especially if the latter dominates the energy balance of the cloud, results in a power-law tail (PL-tail) distribution at the high density part \citep{Klessen_00, VS_ea_07, KNW_11, Collins_ea_12, Girichidis_ea_14}. \footnote{Macroscopic equilibrium in the condition of developed turbulence and gravity means that the motion of the fluid elements is locally completely chaotic, if we neglect the accretion motion driven by gravity. Moreover, the time-averaged chaotic motion has a locally  constant intensity. In other words, the medium is locally homogeneous and isotropic in terms of the chaotic motion. The assumption for steady state requires that each fluid element that moves inwards, towards the centers of collapse, is replaced by another one which has the same density and chaotic motion intensity. The remark of the local character of these properties  is necessary because: first, the turbulence itself is a fractal phenomenon so the intensity of the chaotic motion dependents on the scale and, second, due to the action of gravity the medium thickens more and more as the centers of collapse are approached, the balance of the gravitational and the kinetic energy changes and, as a result, the intensity of the chaotic motion changes too. The presence of an interval of inertial scales yields not only local macroscopic equilibrium, but also a turbulent kinetic energy cascade, i.e. energy is transferred from scale $L$ to the lower scale $L+dL$ and so on till the lowest scale, namely the dissipation scale, is reached, not directly from the highest to the lowest scale.}

Indeed, our assumptions are idealized to some extent since, as it is well known, turbulence is an intermittent phenomenon \citep{Elme_Scalo_04, HF_12} which means that the ratio of the spatial and the temporal fluctuations of the turbulent velocity field to the average velocity increases as the scale is decreased \citep{HF_12}. Besides, turbulence is characterized by large correlation lengths and long correlation times of the turbulent velocity field \citep{Elme_Scalo_04} which violate the macroscopic statistical equilibrium. That is why we clarified that the medium is homogeneous and isotropic only locally, in small enough regions which, however, are significantly larger than the dissipation scale.

\subsection{Geometric setup of the model in the context of (the concept of) ensemble of MCs}
\label{subsec-model geometry}

Let us denote the volume weighted distribution of the mass density $\rho$ by $p(\rho)$ (Here the time averaged density distribution is assumed.). Let $\rho$ vary in the interval $[\rho_{\rm c},\rho_{0}]$, where $\rho_{0}\gg \rho_{\rm c}$. The corresponding logarithmic density $s\equiv \ln(\rho/\rho_{\rm c})$ takes values in $[0,s_{0}]$. It is more convenient to work with the distribution  of the logarithmic density $p(s)$ instead of $p(\rho)$, because $s$ is the natural variable when the density varies within several orders of magnitude. Besides, $p(\rho)$ can be easily found from $p(s)$ through the relation $p(\rho)=p(s(\rho))$. From now on we will term $p(s)$ probability distribution function (PDF) of the density, and $s$  -- simply density.

The characteristic scale of our problem is $l_{\rm c}$. Then, the scale $L(s)$ corresponding to density $s$, is:
\begin{equation}
\label{def_scale}
L(s)=l_{\rm c}\bigg(\int\limits_{s}^{s_{0}}p(s')ds'\bigg)^{1/3}~.
\end{equation}
This definition conforms with $p(s)$ as the latter is volume weighted, i.e:
\begin{equation}
\label{def_p}
p(s)ds\equiv\frac{dV(s)}{V_{\rm c}}~,
\end{equation}

where $V_{\rm c}\propto l_{\rm c}^3$ is the total volume of the system (cloud), and $dV(s)$ is the volume in which the density is in $[s,s+ds]$.

Hence, we arrive naturally to the idea that our system, the molecular cloud, can be related to an abstract gas ball the density of which increases monotonously from its boundary ($\rho=\rho_{\rm c}$, $s=s_{\rm c}$) to the center ($\rho=\rho_{0}$, $s=s_{0}$) in such a way that the PDF of the two systems is the same. The scale $L(s)$ is, actually, the radius of the embedded sphere at which the local density takes the value $s$.
The volumes $dV(s)$ are thin shells with radii lying in $[L(s),L(s+ds)]$. The abstract object that we describe, the gas ball, can be thought of as a typical representative of a class that contains all the physical systems, MCs, which have the same $p(s)$, $l_{\rm c}$ and $T$ \citep{DVK_17}. The turbulence in this abstract object is locally isotropic and homogeneous in each shell $dV(s)$.

A similar geometric setup is proposed by \citet{LB_16} (see Fig.1 in their paper). They relate the column density PDF obtained from observations to a ball-shaped abstract molecular cloud. The radii of its embedded spheres correspond to scales obtained through a formula much similar to the one that we use (\ref{def_scale}). The difference between their and our model is that they do this for the specific case of a power-law PDF, while we pose no constraints on the particular function. Moreover, they treat their model as a simplified description of a given, real physical object, while for us the abstract ball is treated statistically as the typical representative of an ensemble or a class of real objects.

Under what condition can we have real statistical equilibrium and, ergo, non-varying $p(s)$? In the real system, gravity causes matter to fall towards the centers of the clumps of higher density. It means that in the abstract object, with which we will work from now on, the matter falls towards the center of the gas ball. If the gas ball was a closed system, however, its PDF would vary with time. This problem can be resolved if matter which has density $\rho_{\rm c}$, turbulent velocity fluctuations of magnitude corresponding to the scale $l_{\rm c}$ and some initial accretion velocity,  is continuously supplied through the outer boundary of the system (see the bathtub model in \citet{Burkert_17}). On the other hand, the matter which falls on the center accumulates and forms a small but very dense core the influence of which will be taken into account when the gravitational potential is evaluated below. The high density of the core justifies the assumption that these quantities remain almost unchanged in time intervals relatively short in comparison to the lifetime of the MC. This, to some extent idealized, setup guarantees statistical equilibrium.

The latter assumption requires that the radius of the core $l_0$ should be formally added in (\ref{def_scale}). However, we are interested in the inertial region, so we will assume that $L(s)\gg l_0$ which allows us to neglect the radius of the core in the formulas that follow.

One more clarification concerning the dynamics of our abstract object is due here. As it is spherically symmetric and the chaotic turbulent motion in each shell is isotropic and homogeneous, the equations of the medium will have the simplest possible form. It is clear that the dynamics that they represent does not correspond completely to the dynamics of each of the MCs in the ensemble. We hope, however, that the major processes will be reproduced as our abstract object is a direct and natural consequence of the PDF of the mass density which is a major characteristic of all the members of the MC ensemble. The PDF contains much information about the structure and, hence, the dynamics of the clouds.

\section{Derivation of the equations for the PDF of the mass density}   \label{Sec-Equ_p(s)}

\subsection{Derivation of the general form of the equations of the medium}
\label{subsec-equ_p(s)_general}

Our task is to obtain an equation which determines the distribution $p(s)$ within the model that we have build above. We start form the equations of the medium. They are:

- The system of compressible Euler equations

\begin{equation}
\label{equ_N-St}
\rho \frac{\partial\vec{u}}{\partial t} + \rho \vec{u}\cdotp\nabla\vec{u} = -\nabla p_{\rm th} - \rho\nabla\varphi
\end{equation}

\begin{equation}
\label{equ_cont}
\frac{\partial\rho}{\partial t} + \nabla\cdotp(\rho\vec{u})=0
\end{equation}

- The equation of state of the gas (\ref{Cl-M_cs}), which we have given already but remind it here

\begin{equation}
\label{equ_id-gas}
p_{\rm th}=c_{\rm s}^2 \rho
\end{equation}

- The Poisson equation for the gravitational potential

\begin{equation}
\label{equ_Pois}
\Delta\varphi=4\pi G \rho~.
\end{equation}

As it can be seen, in the equation (\ref{equ_N-St}), which is, actually, the equation of motion of a fluid element,  the external force which introduces kinetic energy in the medium and the dissipative terms have been neglected. This is justified due to the statistical equilibrium which is characteristic for the inertial interval of scales. The equation of continuity (\ref{equ_cont}), which is actually a differential form of the law of conservation of mass, will later allow us to obtain a scaling law for the accretion velocity -- the velocity at which the matter falls towards the center of the ball.

The equation of state of the ideal gas (\ref{equ_id-gas}) reflects the assumption that the medium is isothermal. The Poisson equation (\ref{equ_Pois}) determines the gravitational potential produced by the given density distribution.

Other factors, such as the angular momentum of the core, the magnetic fields, the back reaction of the newborn stars, have been deliberately neglected, so that the model is not too cumbersome for a first step towards an equation for $p(s)$ to be made.

In what way can these equations be useful? Since we deal with the abstract object, our gas ball, the equation for $p(s)$ must be written in form which takes into account its symmetries and the physical assumptions for it.

Let us consider the following steps:
\begin{itemize}
\item Let us divide equation  (\ref{equ_N-St}) in the density $\rho$

$$ \frac{\partial\vec{u}}{\partial t} + \vec{u}\cdotp\nabla\vec{u} = -\frac{1}{\rho}\nabla p_{\rm th} - \nabla\varphi,$$

\item From (\ref{equ_id-gas}) it follows that $\nabla p_{\rm th}=c_{\rm s}^2 \nabla \rho$. Since $\nabla\rho/\rho = \nabla \ln(\rho/\rho_{\rm c}) = \nabla s$, it follows that

$$\frac{\partial\vec{u}}{\partial t} + \vec{u}\cdotp\nabla\vec{u} = -c_{\rm s}^2\nabla s - \nabla\varphi,$$

\item Multiplication of the above expression by an infinitesimal displacement   $d\vec{r}=\vec{u}dt$ in the direction of the vector field $\vec{u}$ gives

$$\vec{u}\cdotp\frac{\partial\vec{u}}{\partial t}dt + (\vec{u}\cdotp\nabla\vec{u})\cdotp\vec{u}dt = -c_{\rm s}^2(\nabla s)\cdotp d\vec{r} - (\nabla\varphi)\cdotp d\vec{r},$$

\item It can be easily found that

$$\vec{u}\cdotp\frac{\partial\vec{u}}{\partial t}dt + (\vec{u}\cdotp\nabla\vec{u})\cdotp\vec{u}dt = \frac{d}{dt}(u^2/2)dt = d(u^2/2),$$

\item If the dimensionless variables $v^2\equiv u^2/c_{\rm s}^2$  and $\phi\equiv\varphi/c_{\rm s}^2$ are introduced, then the equation takes the form

$$d(v^2/2) = -(\nabla s)\cdotp d\vec{r} - (\nabla \phi)\cdotp d\vec{r},$$

\item On the other side $ds = (\partial s/\partial t)dt + (\nabla s)\cdotp d\vec{r}$ and $d\phi = (\partial\phi/\partial t)dt + (\nabla\phi)\cdotp d\vec{r}$, so

$$d(v^2/2) = -(ds - (\partial s/\partial t)dt) - (d\phi - (\partial\phi/\partial t)dt)~.$$
\end{itemize}

From now on we assume that all the quantities pertain to the dynamics of our abstract object. The above considerations were more general and did not require such assumption to be made. In order to stress that the PDF of the mass density has been obtained through statistical averaging we put the quantities in brackets. On the one side, this notation expresses the idea for the design of the abstract object. On the other side, we assume ensemble averaging\footnote{The averaging is not over the ensemble of molecular clouds with same PDFs. It is, rather, over the ensemble of the microscopic states which realize the macroscopic system -- the abstract representative of the ensemble of MCs.} with respect to the dynamics of the fluid elements in each shell. We can write:

$$d(\langle v^2/2 \rangle) = -(d\langle s \rangle - (\partial \langle s \rangle/\partial t)dt) - (d\langle \phi \rangle - (\partial \langle \phi \rangle/\partial t)dt)~.$$

At this point we take advantage of the assumption that the system is in steady state hence $\partial \langle s \rangle/\partial t = 0$ and $\partial \langle \phi \rangle/\partial t = 0$. The equation takes the form:

\begin{equation}
\label{dE=0}
d[\langle v^2/2 \rangle + \langle s \rangle + \langle \phi \rangle]=0~.
\end{equation}
We should not forget that the total differential here corresponds to a shift in the direction of motion of the fluid element. The ensemble averaged motion (in the above mention sense) is radial fall towards the center of the ball. The turbulence is locally homogeneous and isotropic so its contribution to the average motion vanishes. Since the object is spherically symmetric there is one to one correspondence between radial position and scale. Then, the total differential expresses not only a radial shift but also a shift along the scales $L(s)$. Within this interpretation equation (\ref{dE=0}) expresses the conservation of the energy per unit mass of the fluid element as is passes from scale $L$ to scale $L+dL$. The fact that  $L(s)$ is monotonous allows us to write equation (\ref{dE=0}) in the following equivalent form:
\begin{equation}
\label{dE/ds=0}
\frac{d}{ds}[\langle v^2/2 \rangle + \langle s \rangle + \langle \phi \rangle]=0~.
\end{equation}
We should clarify that after the averaging the turbulent velocity gives no contribution to the motion of the fluid element. It gives, however, non vanishing contribution to the kinetic energy term  $\langle v^2/2 \rangle$ due to the scalar nature of the latter. (The velocity is first squared then averaged.)

\subsection{Explicit form of the terms in (\ref{dE/ds=0})}
\label{subsec-equ_p(s)_terms}
In this subsection we derive the explicit form of the terms in equation (\ref{dE/ds=0}) taking into account the model presented in Section \ref{Sec-Model}. Let us start with the kinetic energy term which can be expressed as:
\begin{equation}
\label{v=vt+va}
\langle v^2 \rangle = \langle v_{\rm t}^2 \rangle + \langle v_{\rm a}^2 \rangle~,
\end{equation}
where $\langle v_{\rm t}^2 \rangle$ is the kinetic energy per unit mass coming form the turbulent velocity fluctuations, and $\langle v_{\rm a}^2 \rangle$ is the kinetic energy per unit mass coming form the matter accretion towards the center of the cloud. To prove that (\ref{v=vt+va}) is satisfied we decompose the velocity in the following way:
$$ \vec{v} = \vec{v}_{\rm t} + \vec{v}_{\rm a}~\Rightarrow~ v^2 = v_{\rm t}^2 + v_{\rm a}^2 + 2\vec{v}_{\rm t}\cdotp\vec{v}_{\rm a}~, $$
then after the averaging we obtain
$$\langle v^2 \rangle = \langle v_{\rm t}^2 \rangle + \langle v_{\rm a}^2 \rangle + 2\langle \vec{v}_{\rm t}\cdotp\vec{v}_{\rm a} \rangle~.$$

Then, if we assume that the two velocity components are independent, $\langle \vec{v}_{\rm t}\cdotp\vec{v}_{\rm a} \rangle = \langle \vec{v}_{\rm t} \rangle\cdotp\langle \vec{v}_{\rm a} \rangle = 0$, since $\langle \vec{v}_{\rm t} \rangle = 0$. The latter is a consequence of the fact that the turbulence is locally homogeneous and isotropic. Even if we assume that the accretion and the turbulence components are not independent, as in \citep{KH_10}, the last term in $\langle v^2 \rangle$ still vanishes because the turbulent component is chaotic while the accretion component is purely radial.

Since our spherically symmetric cloud is ensemble averaged we could apply a standard scaling relation for  $\langle v_{\rm t}^2 \rangle$, which, in addition, reflects the fractal structure of our object \citep{Kritsuk_ea_07}:
\begin{eqnarray}
\label{vt2-p(s)}
\langle v_{\rm t}^2 \rangle = \frac{u_0^2}{c_{\rm s}^2}\bigg(\frac{l_{\rm c}}{pc}\bigg)^{2\beta} \bigg(\frac{L(s)}{l_{\rm c}}\bigg)^{2\beta} = \nonumber \\ \frac{u_0^2}{c_{\rm s}^2}\bigg(\frac{l_{\rm c}}{pc}\bigg)^{2\beta}\bigg(\int\limits_{s}^{s_{0}}p(s')ds'\bigg)^{2\beta/3}~,
\end{eqnarray}
where $u_0$ and $\beta$ are, respectively, the normalizing factor and the scaling exponent of the  turbulent velocity fluctuations in the standard law $u=u_0 L^{\beta}$ \citep{Larson_81, Pad_ea_06}. We chose to use the relation between the accretion and the turbulent kinetic energy (for given values of the density of the medium), that has been ascertained by \citet{KH_10} because it makes the model more self-consistent and decreases the number of free parameters in the same time. According to their result and our assumption of dynamical equilibrium in each shell of the cloud:
\begin{equation}
\label{vt2-va2}
\langle v_{\rm t}^2 \rangle = \exp(-s) \langle v_{\rm a}^2 \rangle~.
\end{equation}
The explicit form of the accretion term of the kinetic energy will be given in the following subsection, as it requires a subtle analysis of equation (\ref{equ_cont}).

We are well aware that the mathematical simplicity of the equations is not a strong enough argument for the replacement of the standard scaling law of the turbulent energy (\ref{vt2-p(s)}) with equation (\ref{vt2-va2}). This replacement would change the physics woven into the model and would alter the conclusions that could be drawn from the presence of particular types of solutions of the equation for the density distribution (\ref{E_Q}).  Actually the use of (\ref{vt2-va2}) means we assume that turbulence is driven by accretion which, on its side, originates at scales much greater than the size of the molecular cloud and persists inside it due to its self gravity.
This, namely, is the idea of the authors of \citet{KH_10}. In a broader sense by this replacement we make our model to partially conform with the thesis of several authors form the last years \citep{Heyer_ea_2009, BP_ea_2011a, BP_ea_2011b, Traficante_ea_2015, I-M_ea_2016} according to which non-thermal velocity dispersion (at the level of both galactic interstellar medium and molecular clouds) is a manifestation of hierarchical, chaotic gravitational collapse. Though supported by a number of observations and simulations in the cited works this thesis still bounds us to a given paradigm. A more unbiased study of the role that turbulence has in the energy balance would be based on equation (\ref{vt2-p(s)}). This, however, involves the study of a different equation (see equation (\ref{E_vt2_Q}) in Section \ref{subsec-model divelopment}) which is a task on its own that we leave for a future work.

The thermal potential  $\langle s \rangle$ is simply equal to $s$, since in our model this is how the average logarithmic density is denoted. The same is valid for the density itself: $\rho=\langle\rho\rangle$.

The averaged gravitational potential, due to the spherical symmetry of the cloud, is simply:
\begin{equation}
\label{grav-potencial}
\langle \phi \rangle= -\frac{G}{c_{\rm s}^2}\frac{M(s)}{L(s)} -\frac{G}{c_{\rm s}^2}\frac{M_0}{L(s)}~,
\end{equation}
where $L(s)$ is the radius at which the fluid element resides at the given moment, and $M(s)=M_{\rm c}^{*}\int_{s}^{s_0} \exp(s')p(s')ds'$ is the mass corresponding to $L(s)$ (the mass of the core is not included). Both of these quantities, and hence  $\langle \phi \rangle$ are expressed by the distribution $p(s)$:
\begin{eqnarray}
\label{grav-potencial_p(s)}
\langle \phi \rangle= -\frac{G}{c_{\rm s}^2}\frac{M_{\rm c}^{*}}{l_{\rm c}} \frac{\int\limits_{s}^{s_0} \exp(s')p(s')ds'}{\bigg(\int\limits_{s}^{s_0}p(s')ds'\bigg)^{1/3}} \nonumber \\ -\frac{G}{c_{\rm s}^2} \frac{M_0}{l_{\rm c}} \frac{1}{\bigg(\int\limits_{s}^{s_0}p(s')ds'\bigg)^{1/3}}~,
\end{eqnarray}
where $M_{\rm c}^{*}= (4/3)\pi l_{\rm c}^3 \rho_{\rm c}$ is a normalizing coefficient the physical interpretation of which is rather clear -- this is the mass of a ball with radius equal to the outer radius of the cloud and density equal to that of the boundary of the cloud. (Its mass is lower than that of the cloud by a factor of few.) $M_0$ is the mass of the dense core at the center of the cloud.

As it can be seen, in the expression for $\langle \phi \rangle$ only the influence of the mass contained inside the ball of radius $L(s)$ is taken into account. The masses of the outer shells also contribute to the gravitational potential but exert no force on the fluid element located at distance $L(s)$ from the center of the ball. We obtained  equation (\ref{dE/ds=0}) starting from the equation of motion (\ref{equ_N-St}), the right-hand side of which contains the gravitational force $-\nabla\varphi$, not the total gravitational potential. With this in mind, we have to include only those components of the gravitational potential $\langle \phi \rangle$ that produce force.

If we split the solution of Poisson equation  (\ref{equ_Pois}) in two parts, one representing the contribution of the mass contained in a sphere with radius $L(s)$ and another one coming from the outer (with respect to $L(s)$) layers, then $\langle \phi \rangle$ in equation (\ref{grav-potencial_p(s)}) coincides with the former.    

\subsection{The continuity equation and the explicit form of $\langle v_{\rm a}^2 \rangle$}
\label{subsec-equ_p(s)_accretion term}

Let us average equation (\ref{equ_cont}) with respect to the ensemble of the microstates of the abstract object (just as we did in subsection \ref{subsec-equ_p(s)_general}). We have:
$$\frac{\partial\rho}{\partial t} + \nabla\cdotp\langle\rho\vec{u}\rangle=0~.$$
Due to the fact that the system is in steady state: $\partial\rho/\partial t =0$. Then:
$$\nabla\cdotp\langle\rho\vec{u}\rangle=0~.$$
In the natural for our model spherical coordinates, the velocity of the fluid element is: $\vec{u}=(u_{L},u_{\vartheta},u_{\varphi})$. So the momentum per unit volume is:  $\rho\vec{u}=(\rho u_{L},\rho u_{\vartheta},\rho u_{\varphi})$. As mentioned above in subsection \ref{subsec-equ_p(s)_terms}, the velocity has two components $ \vec{u}_{\rm t}$ and $\vec{u}_{\rm a}$, turbulent and accretion, respectively. The first one is completely chaotic, while the second one points towards the center. Hence, the momentum also has two components : $\rho\vec{u} = \rho\vec{u}_{\rm t} + \rho\vec{u}_{\rm a}$. Then, after ensemble averaging $\langle\rho\vec{u}\rangle = \langle\rho\vec{u}_{\rm t}\rangle + \langle\rho\vec{u}_{\rm a}\rangle = \rho\langle\vec{u}_{\rm t}\rangle + \rho\langle\vec{u}_{\rm a}\rangle = \rho\langle\vec{u}_{\rm a}\rangle$. Hence, because $\langle\vec{u}_{\rm a}\rangle = (\langle u_{L} \rangle, 0, 0)$, in spherical coordinates the continuity equation takes the form:
$$\nabla\cdotp(\rho\langle u_{L} \rangle,0,0)=0~~\Leftrightarrow~~ \frac{1}{L^2}\frac{\partial}{\partial L}(L^2\rho\langle u_{L} \rangle)=0.$$
It is clear that $\rho$ and $\langle u_{L} \rangle$ depend only on the radius $L=L(s)$ and since $\langle u_{\rm a}\rangle = \langle u_{L} \rangle$, we arrive at the following equation:
$$L^2\rho\langle u_{\rm a} \rangle=const(L)~.$$
Taking into account that the average density at level $L(s)$ is  $\rho=\rho_{\rm c}\exp(s)$, we obtain for the dimensionless velocity    $\langle v_{\rm a}\rangle\propto \exp(-s)(L/l_{\rm c})^{-2} = \exp(-s)\bigg(\int\limits_{s}^{s_{0}}p(s')ds'\bigg)^{-2/3}$. The accretion velocity has only radial component, which is not chaotic, so $\langle v_{\rm a}^2\rangle= \langle v_{L}^2\rangle= \langle v_{L}\rangle^2 =\langle v_{\rm a}\rangle^2$ and hence we arrive at the following expression for the accretion term:
\begin{equation}
\label{va2-p(s)}
\langle v_{\rm a}^2\rangle\propto \exp(-2s)\bigg(\int\limits_{s}^{s_{0}}p(s')ds'\bigg)^{-4/3}~.
\end{equation}
The dimensionless coefficient that we will introduce shortly in order to convert (\ref{va2-p(s)}) to equality, is, actually, the ratio of the accretion kinetic energy term at the boundary of the cloud and the thermal kinetic energy per unit mass  (the latter is simply equal to the square of the speed of sound).

\subsection{Derivation of the equation for $p(s)$}
\label{subsec-equ_p(s)_diff equ}

In order to simplify the calculations we introduce the following variable:
\begin{equation}
\label{Q-def}
Q(s)\equiv \bigg(\int\limits_{s}^{s_{0}}p(s')ds'\bigg)^{1/3}~.
\end{equation}
The function $Q(s)$ has physical interpretation -- it is the third root of the probability, evaluated by the distribution $p(s)$ up to the level $s$ or, which is even better, according to  (\ref{def_scale}) it is the dimensionless radius. With this variable the  terms of equation (\ref{dE/ds=0}) can take their dimensionless form. According to equation (\ref{va2-p(s)}) we have $\langle v_{\rm a}^2\rangle\propto \exp(-2s) Q^{-4}(s)$. Beqause of equations (\ref{vt2-va2}) and (\ref{va2-p(s)}) it folows $\langle v_{\rm t}^2\rangle\propto \exp(-3s) Q^{-4}(s)$. For the gravitational potential $\langle\phi\rangle$ we have two terms. According to equation (\ref{grav-potencial_p(s)}) the first one is $\propto -\int\limits_{s}^{s_0} \exp(s')p(s')ds'/Q(s)$. Taking advantage of the facts that $Q(s_0)=0$ and $p(s)=-d(Q^{3}(s))$ we reorganize the integral which appears in the denominator in the following way:
\begin{eqnarray}
\label{int_Q-transform}
\int\limits_{s}^{s_0} \exp(s')p(s')ds'= \int\limits_{s}^{s_0} \exp(s')d(-Q^3(s'))= \nonumber \\ -(\exp(s_0)Q^3(s_0)-\exp(s)Q^3(s)) \nonumber \\
+ \int\limits_{s}^{s_0} \exp(s')Q^3(s')ds'= \nonumber \\
 \exp(s)Q^3(s) + \int\limits_{s}^{s_0} \exp(s')Q^3(s') ds'~.
\end{eqnarray}
So for the gravitational potential contribution of the inner shells (without the core) we obtain: $-(\exp(s)Q^3(s) + \int_{s}^{s_0} \exp(s')Q^3(s') ds')/Q(s)$. The gravitational potential that comes only from the core, obviously is $\propto -1/Q(s)$. Finaly, accounting for dimensionless coefficients, the terms of equation (\ref{dE/ds=0}) reads:

\begin{eqnarray}
\label{vt2_va2_phi-Q}
\langle v_{\rm a}^2\rangle= A_0 \exp(-2s)Q^{-4}(s)~,\nonumber \\
\langle v_{\rm t}^2\rangle= A_0 \exp(-3s)Q^{-4}(s)~,\nonumber \\
\langle\phi\rangle=- G_0 \frac{\exp(s)Q^3(s) + \int_{s}^{s_0} \exp(s')Q^3(s') ds'}{Q(s)} \nonumber \\
- \frac{G_1}{Q(s)}~.
\end{eqnarray}

The dimensionless coefficients in the last formula of equation (\ref{vt2_va2_phi-Q}) are: $G_0=(G/c_{\rm s}^2)(M_{\rm c}^{*}/l_{\rm c})$, $G_1=(G/c_{\rm s}^2)(M_0/l_{\rm c})$. Their physical  interpretations are, respectively, the ratio of the gravitational energy of the entire cloud (without the core)  and its thermal energy per unit mass, and the ratio of the gravitational energy per unit mass of the core and its thermal energy per unit mass. Both of them are evaluated for the outer layer of the cloud. The coefficient $A_0$ has a similar meaning but for the accretion kinetic energy per unit mass and was discussed at the end of the previous subsection.
With this preparation equation (\ref{dE/ds=0}) can be written in the following form:
\begin{eqnarray}
\label{dE/ds=0_Q}
\frac{d}{ds}\Bigg[A_0 (1+\exp(-s)) \exp(-2s)Q^{-4}(s) + s\nonumber \\
- G_0 \frac{\exp(s)Q^3(s) + \int_{s}^{s_0} \exp(s')Q^3(s') ds'}{Q(s)} -\frac{G_1}{Q(s)}\Bigg]=0
\end{eqnarray}
Let us denote the expression in the brakets as $E_0$. This is the total energy per unit mass of the fluid element. Then:
\begin{eqnarray}
\label{E_Q}
A_0 (1+\exp(-s)) \exp(-2s)Q^{-4}(s) + s \nonumber \\
- G_0 \frac{\exp(s)Q^3(s) + \int_{s}^{s_0} \exp(s')Q^3(s') ds'}{Q(s)} -\frac{G_1}{Q(s)}=E_0~.
\end{eqnarray}
Equation (\ref{E_Q}) is a nonlinear integral equation for the function $Q(s)$. A  solution for $Q(s)$ would allow us to find the density distribution:
\begin{equation}
\label{p-Q}
p(s)=-3Q^2(s)Q'(s)~,
\end{equation}
as it can be easily seen from (\ref{Q-def}). Here $Q'=dQ(s)/ds$.

\section{Study of the equation for the PDF}	
\label{Sec-analysis and sol}
As a first step, we search for a solution of the form $Q(s)= B\exp(as)$ which corresponds to a PL-PDF, $p(s)\propto \exp(qs)$ with $q=3a$ (see (\ref{p-Q})). The motivation for this ansatz is the following. The most interesting part of the PDF is its PL-tail where the star formation occurs. Besides, à solution of this type is the simplest possible.  A more general approach would be to ask for a solution in the form of a series of decreasing exponents:
\begin{equation}
\label{Q-series}
Q(s)= B_0 \exp(a_0 s) + B_1 \exp(a_1 s) + B_2 \exp(a_2 s) + ...~,
\end{equation}
where $0>a_0>a_1>a_2>...$, since $0\leq s \leq s_0$ \footnote{This idea was proposed by assoc. prof. Angel Zhivkov in private conversation.}. In fact our approach would give the zeroth order approximation. Making this substitution in (\ref{E_Q}) after some algebra we arrive at:
\begin{eqnarray}
\label{Ealg_Q}
A_0 B^{-4} (1+\exp(-s)) \exp((-2-4a)s) + s \exp(0) \nonumber \\
- G_0 B^{2} \frac{\exp((1+3a)s_0)\exp(-as)+3a\exp((1+2a)s)}{1+3a} \nonumber \\
- G_1 B^{-1} \exp(-as) \nonumber \\
=E_0 \exp(0)~.
\end{eqnarray}
We will study the following two cases. In the first case, the core can be neglected, i.e. $A_0, G_0 \gg G_1$. In the second case, the core has a significant contribution or $A_0, G_0 \sim G_1$.

In order to estimate the normalizing coefficient $B$ we use the initial condition at $s=0$, namely $B \exp(0)\approx (\int_{0}^{s_0} p(s')ds')^{1/3}=1$. Hence, $B\approx 1$.

\subsection{Solution in the absence of a core}
\label{subsec-without G1}
When the core is neglected (\ref{Ealg_Q}) takes the form:
\begin{eqnarray}
\label{Ealg_Q_wotG1}
A_0 (1+\exp(-s)) \exp((-2-4a)s) + s \exp(0) \nonumber \\
- G_0 \frac{\exp((1+3a)s_0)\exp(-as)+3a\exp((1+2a)s)}{1+3a}  \nonumber \\
= E_0 \exp(0)~.
\end{eqnarray}
Before we continue let us clarify the physical interpretation of the terms that appear in the above equation. As it can be seen from equation (\ref{vt2_va2_phi-Q}) the first term gives the contribution of the accretion and the turbulent energy. The second term represents the thermal energy (see equation (\ref{dE/ds=0}) and the paragraph above formula (\ref{grav-potencial})). The last term on the lhs is the gravitational potential energy (without the core) explicitly given in equation (\ref{vt2_va2_phi-Q}). The total energy is the only term on the rhs. The exponents of this terms, through our ansatz, are respectively $-2-4a,~-3-4a,~0,~-a,~1+2a,~0$ (note that there are two exponents for the gravitational term).
The comparison of these exponents (comparing each-other) gives the following roots for $a$:  $-1,~-3/4,~-2/3,~-1/2,~-1/3,~0$ \citep{Zhivkov_99, RHB_2006}. The equation is satisfied only for $a=-1/2$. With this value for $a$ the exponents of the corresponding terms are $0~,-1,~0,~1/2,~0,~0$. The term $\exp(s/2)$, which should be of the leading order is diminished by the $\exp(-s_0/2)$ multiplier \footnote{To neglect the core means that we should be far enough from it, i.e. $s\ll s_0~\Rightarrow~\exp((s-s_0)/2)\ll 1$.}. For the same reason $a=-2/3$ is also not a solution, even though it appears at first that the accretion and the gravitational term balance each-other. So for the only root far from the core $a=-1/2$ we have balance between the accretion kinetic energy, the thermal energy and the gravitational energy, because all these terms have exponents of the leading order $0$. The total energy exponent is also of the same order, while the turbulent kinetic energy exponent is negligible, its order is $-1$. Since $A_0, G_0\gg s$ in the considered interval of values of $s$, as we will see from the numerical estimation in subsection \ref{subsec-calc A_0,G_0,G_1}, the accretion and the gravitational energy determine the total energy:
\begin{equation}
\label{E_0}
E_0\approx A_0-3G_0~.
\end{equation}
The equality is approximate as this is the zeroth order approximation of our solution, but still, it can serve as an estimate for the balance of the energy components. Besides, if we take into account the definition of $G_0$ and the average density of the cloud which in the case of a long PL-tail can be evaluated from
$\langle\rho\rangle_{\rm c}= (q/(1+q))\rho_{\rm c}$ $~$ \citep{DVK_17}, which in the case $q=-1.5$ ($a=-1/2$) gives $\langle\rho\rangle_{\rm c}= 3\rho_{\rm c}$, we can obtain the average gravitational energy per unit mass of the fluid element (the averaging is over the entire cloud)  $\langle G \rangle= 3G_0$. For $q=-1.5$ the accretion energy per unit mass remains constant from one scale to the other (see subsection \ref{subsec-results}). Then its average value is $\langle A \rangle= A_0$. Equation (\ref{E_0}) reduces to  $\langle A\rangle - \langle G \rangle \approx E_0$, which is rather natural.

When the core is neglected  the PDF is: $p(s)\approx (3/2) \exp(-3s/2)$.

\subsection{Solution in the presence of a core}
\label{subsec-with G1}
When the core is not negligible equation (\ref{Ealg_Q}) reads:
\begin{eqnarray}
\label{Ealg_Q_wtG1}
A_0 (1+\exp(-s)) \exp((-2-4a)s) + s \exp(0) \nonumber \\
- G_0\frac{\exp((1+3a)s_0)\exp(-as)+3a\exp((1+2a)s)}{1+3a} \nonumber \\
- G_1 \exp(-as) \nonumber \\
= E_0 \exp(0)~.
\end{eqnarray}
The power exponents are: $-2-4a,~-3-4a,~0,~-a,~1+2a,~-a,~0$. The comparison of all the possible pairs gives the same set of roots for $a$: $-1,~-3/4,~-2/3,~-1/2,~-1/3,~0$ \citep{Zhivkov_99, RHB_2006}. Among them, balance between the different terms of the equations is realized only for $-2/3$ and $-1/3$. In the first case the accretions balances the gravitation of the core which leads to: $A_0\approx G_1$. Here the gravitation of the shells above the core is negligible since $s\lesssim s_0$. In the second case ($a=-1/3$), the two gravitational terms balance each-other. This is a rather peculiar from a physical point of view fact. As it can be seen, the coefficient $3a/(1+3a)$ coming from the integral in (\ref{E_Q}) diverges. The integral should be properly evaluated for $Q(s)= B \exp(-s/3)$. When this is done one arrives at: $G_0(1+s_0-s) + G_1 =0$. Besides the unwanted presence of $s$ one notices that the equation requires $s>s_0+1$ which would mean that the fluid element has entered inside the core. As we stated in subsection \ref{subsec-equ_p(s)_terms}, when the fluid element is inside the core the outer layers exert no force on the fluid element. So as it appears the $a=-1/3$ solution must be taken out of  consideration.
At the end, in the presence of a core we have only one solution $a=-2/3$. The PDF that corresponds to it is: $p(s)\approx 2 \exp(-2s)$.

\subsection{Estimate of the numerical values of the coefficients}
\label{subsec-calc A_0,G_0,G_1}

For concreteness we will estimate the dimensionless coefficients. As it can be seen from \cite{KH_10} and \cite{HF_12}, $7~ {\rm km/s}$  is a reasonable lower bound on the velocity at which matter accretes onto the outer boundary of the cloud.  A gas temperature of $10~{\rm K}$ implies that the speed of sound is $c\approx 0.2~ {\rm km/s}$. Hence, $A_0\sim 10^3$. In order to estimate $G_0$ we use the following reasonable values of the physical parameters of the cloud $\rho_{\rm c}\sim 10^3$, $l_{\rm c}\approx 5~{\rm pc}$. As a result, we obtain $G_0\sim 10^3$. Finally, for the estimate of $G_1$ we assume that $\rho_0/\rho_{\rm c}\sim 10^6$ \citep{KNW_11} and that the slope of the density profile (the scale-density relation) is $-2$, which means that $a=-1/2$, and obtain $G_1\sim 1$. With these assumptions, which are justified for a large number of observed and simulated objects, even the stronger condition $A_0, G_0 \gg G_1\exp(-as)$ for $-as<7$ holds. The latter means that the core can be neglected when $s<14$ or $\rho_0/\rho_{\rm c}\sim 10^6$, which is in agreement with our assumption. The latter is of considerable importance in the light of the results of subsection \ref{subsec-without G1}. Finally, we would like to note that if the density profile was flatter, i.e. for steeper  $Q(s)$ and $p(s)$,  the strong inequality $A_0, G_0 \gg G_1\exp(-as)$ would be true even for lower density contrasts observed at earlier stages of the development of the PL-tail.

When  $a=-2/3$, or $q=-2$ (i.e. the density profile is $-1.5$), we have a solution for which  $A_0\approx G_1$. This means that the density contrast is low. The core is close to the outer boundary of the cloud. The span of such a tail would be within a decade of densities (we are talking about  $\rho$ here). It appears that the second solution is a continuation of the first one in the region close to the core. The $a=-2/3$ solution describes the free fall of matter towards the core, while the $a=-1/2$ solution corresponds to a dynamical equilibrium in the outer layers of the cloud  between the accretion kinetic and the gravitational terms in the equation.

\section{Discussion}
\label{Sec-discussion}

In this section we discuss our assumptions and results and compare the latter to similar results in the field. At the end, we propose directions for further development.

\subsection{Assumptions and basis of the model}
\label{subsec-basic model assump}
Our model has three basic assumptions. First, we represent the MC as a spherically symmetric object with radii corresponding to the abstract scales which are naturally related to the PDF (see equation \ref{def_scale}). We treat this object as the average member of the class of objects which have the same PDF $p(s)$, (maximum) size $l_{\rm c}$ and temperature $T$. We should stress that the abstract object is not meant to reproduce the morphology of the different members of the class which can be very diverse. Second, we assume that it is in microscopic (in terms of the thermal motion) and macroscopic (in terms of the motion of the fluid elements) equilibrium in each scale, so the turbulence is developed, locally homogeneous and isotropic. Hence, $p(s)$ is time invariant. In other words, we have a micro- and  macroscopic steady state. The scales of the cloud are, according to our consideration, part of the inertial interval of the developed supersonic turbulence. Finally, we assume that the system is isothermal.

The last two assumptions are idealizations, or rather, a first order approximation of the complex objects that we are dealing with. In this sense the model can be further elaborated. We are fully aware of the fact that the fluid dynamics and thermodynamics of our spherically symmetric cloud does not reproduce the whole diversity and complexity met in the class. We believe, however, that it encompasses the main elements of their physics. We treat the first assumption not as a simplification but rather as an attempt for generalization. The introduction of the  concept of an abstract average object which represents an entire class of similar objects unified by their PDF of the mass density \citep{DVK_17} is an attempt to treat MCs from a statistical point of view.

\subsection{Basic results}
\label{subsec-results}

Starting from the assumptions outlined above we write the equations for the medium  (\ref{equ_N-St}, \ref{equ_cont}, \ref{equ_id-gas} and \ref{equ_Pois}) neglecting the forces which introduce energy into the molecular cloud at large scales and the dissipative terms in the Naiver-Stokes equation since the MC scales belong to the inertial interval. Assuming steady state and averaging over the ensemble of microstates of the abstract object  from equation (\ref{equ_N-St}) we obtain an equation for the conservation of energy per unit mass of a fluid element during its motion through the scales of the MC (\ref{dE/ds=0}). This is the basic result stemming from the theoretical treatment of the model. Another one is the derivation of the formula for the accretion velocity (\ref{va2-p(s)}) through the averaging of the  equation of continuity. Substitution with the explicit form of the energies brings us to a nonlinear integral equation for the function $Q(s)$ (\ref{E_Q}).

The study and the solution of equation (\ref{E_Q}) is tough. We leave it for future work. The initial analysis done in section \ref{Sec-analysis and sol} reveals that a solution close to a PL-PDF, corresponding to a dynamical equilibrium of all the forces in the MC with slope $q=-1.5$ is feasible. For this solution, it is easy to find that the density profile scales as: $ \rho  \propto L^{-2}$ and then, according to (\ref{va2-p(s)}), the accretion velocity is invariant with respect to the scales. The turbulent velocity, however, is proportional to the scale $v_{\rm t}\propto L$, according to (\ref{vt2-va2}). It is easy to obtain that the accretion driven turbulence cascade  satisfies the invariant  $\rho v_{\rm t}^3/L=const(L)$ \citep{Kritsuk_ea_07, HF_12}.
This result deserves special attention as it shows that the accretion driven turbulence scales in a way different from the forced supersonic turbulence with negligible self-gravity which is known to scale as $v_{\rm t}\propto L^{1/2}$. In spate of this different scaling law we see that the energy density flux through the scales is invariant. In order to shed more light on this matter we have to give account of the reason for the presence of the above result. There are two prerequisits for it: first, the particular form of the density profile and the accretion velocity which correspond to the given solution of the equations and, second, relation (\ref{vt2-va2}) which emanates from the simulations made in \citet{KH_10} (the setup of which is rather general). Analytical arguments in support of relation (\ref{vt2-va2}) are also provided in this work. Even though our model is well-founded one might feel more satisfied if the relation between the accretion velocity and the turbulent velocity stems from the equations. This, however, has not been accomplished yet. The natural question that arises here is why this scaling is not observed at least in clouds which are at a late stage of their evolution. Our explanation is based on the paradigm on which (\ref{vt2-va2}) is based. (We have commented it on right after the equation in section \ref{subsec-equ_p(s)_terms}). The essence of this paradigm is that non-thermal velocity dispersion observed in MCs must not be attributed to turbulence but to accretion driven by the self-gravity of the MCs (the collapse is hierarchical and chaotic). According to the particular solution that we are discussing the energy balance of the fluid element is determined by the accretion kinetic energy and gravitational energy. The turbulent kinetic energy has negligible contribution. (The term corresponding to it is of lower order.). In other words, of the two components of the kinetic energy -- accretion and turbulent, it is only the former that has significant contribution and observable effect. Turbulence is unobserved. Can the observed non-thermal velocity dispersion be attributed to  accretion rather than turbulence, then? In order to answer this question, let us apply the condition for the scaling of the non-thermal velocity dispersion $\Delta v \propto \Sigma^{1/2} L^{1/2}$ given in \citet{Heyer_ea_2009, BP_ea_2011a, Traficante_ea_2015, I-M_ea_2016} to our solution for the accretion velocity and check, if it is satisfied. In our case $\Sigma\propto \rho L\propto L^{-2} L\propto L^{-1}~\Rightarrow~\Delta v\propto L^{-1/2} L^{1/2}\propto const(L)$ which is true for the accretion velocity $v_{\rm a}\propto const(L)$ and confirms our thesis. It appears that the model is self consistent since the consequences of the basic solution (basic in sense that it spans the entire density interval far from the core) are consistent with the assumptions leading to (\ref{vt2-va2}).

We also saw that close to the core the equation admits a second PL-PDF solution with slope $q=-2$, which corresponds to free fall of the matter towards the center of the cloud \citep{Shu_77, Hunter_77, Larson_69, Penston_69a}. Here the density profile is $ \rho  \propto L^{-1.5}$, the accretion velocity scales as $v_{\rm a}\propto L^{-0.5}$, which means that it increases as the core is approached, while the power in the scaling law of the turbulent velocity is  positive  but has a lower absolute value: $v_{\rm t}\propto L^{0.25}$. It appears that in this case the cascade turbulence invariant is not satisfied.

When it comes to the contribution of the different terms we can say that  gravity and accretion dominate. For the solution which is valid far from the core,  accretion and gravity of the layers above the core, $A_0$ and $G_0$ respectively, determine the total energy per unit mass of the fluid element: $E_0\approx \langle A \rangle-\langle G \rangle = A_0-3G_0$. For the other solution, the one valid close to the core, we have $A_0\approx G_1$, i.e. there is equipartition of the accretion energy and the gravitational energy produced by the core.

The natural question for the matching of the two solutions cannot be satisfactorily solved at that level of approximation. More terms in the expansion of the solution  (\ref{Q-series}) are necessary.

An important feature of equation (\ref{E_Q}) is the presence of singularity at the right end of the interval. The singularity  in the accretion and the gravitational terms of (\ref{dE/ds=0_Q}) cannot be removed since it is inherent to the model and is related to the neglect of  finite radius of the core $l_0$ still at equation (\ref{def_scale}).

\subsection{Comparison to other theoretical models and simulations}
\label{subsec-comparison}

Similar models have been proposed in the past (analitical as \citet{Penston_69b, Shu_77, Hunter_77}, as well as numerical models such as  \citet{Larson_69, Penston_69a}). They study the collapse of the protostellar  core but the physics and the mathematics involved resemble those of our model (spherical symmetry, isothermal medium, accretion and gravity). The major differences are that turbulence is not included in the consideration and that  accretion is due to the free fall of the matter inside the cloud without matter supply through the boundary of the cloud. Besides, in analitical models, the equations are written directly in spherical coordinates and no averaging is made since in these models the idea of an ensemble of microstates which realize the object is not present  (due to the absence of turbulence).  Exact, self-similar solution, depending on the time and the radial coordinate, are found through an anzats (\citet{Shu_77}). Here, we eliminate the time with the assumption of steady state. The radial density profile, obtained by \citet{Shu_77}, corresponds to the two slopes which we obtain at the level of the leading order term in the approximation: $q=-1.5$ for the outer layers of the cloud (dynamical equilibrium in contrast to static equilibrium in \citet{Shu_77}) and $q=-2$ for the inner layers (free fall). We treat this partial coincidence as a confirmation of the physical and mathematical credibility of our calculations.

From the comparison to simulations \citep{Klessen_00, KNW_11, Collins_ea_12, Girichidis_ea_14, Schneider_ea_14} and observations \citep{Kainu_ea_09,Schneider_ea_14, Schneider_ea_15} the most important confirmation that we would like to stress on is that equation (\ref{E_Q}) has a solution far from the core  which is close to a PL-PDF. This qualitative behavior is of major importance. Its presence is encouraging and makes us believe that our model provides an adequate description of the real objects. There is good agreement in quantitative sense also. The slopes of the tails obtained in the simulations: \citet{KNW_11, Collins_ea_12} are in the interval $-2\leq q \leq-1.5$. Our cloud is in steady state which corresponds to the late stages of evolution. \citet{Girichidis_ea_14} obtain through  theoretical reasoning and numerical calculations that at the final stage of the evolution of the MC  $q=-1.54$, which is rather reassuring.

It is also very important to note that in some modern observations \citep{Schneider_ea_15} and simulations \citep{KNW_11} two tails occur. The slope of the first one is  $-2\leq q \leq-1.5$, while that of the second one is $q\sim-1$ (if, in the interpretation of the observational data, spherical symmetry of the studied MCs is assumed). As the authors of the two cited papers state, the first PL-tail can be explained by the domination of gravity at the corresponding spatial scales. For the second one, they suggest that it is caused by the decrease of the mass flow rate (fall under the action of gravity), from the larger to the smaller substructures of the cloud. This decrease is of unknown origin. It could be caused by a series of factors such as: angular momentum of the small dense structures \citep{KNW_11, Schneider_ea_15}, large opacity and, respectively, pressure increase as a consequence of temperature increase (i.e. the system leaves the isothermal regime), the presence of magnetic fields, the backreaction on the cloud from the newborn stars \citep{Schneider_ea_15} etc. It is natural then that this slope occurs in our model, even though it is not completely correct since the model does not take into account any of the listed factors. This is a hint of the possible directions of elaboration of our model.

\subsection{Directions for future development}
\label{subsec-model divelopment}

Within the frame of our model we see the following two directions of further development of the study. The first one is to obtain a first order differential equation for $Q(s)$ through an elimination of the integral in equation (\ref{E_Q}):
\begin{eqnarray}
\label{diff-equ_Q-1ord}
 - A_0(2+3\exp(-s))\exp(-2s)Q \nonumber \\
   - 3A_0(1+\exp(-s))\exp(-2s)Q' \nonumber \\
 + Q^5 + sQ^4 Q' - 3G_0\exp(s)Q^6 Q' - E_0Q^4 Q' = 0~.
\end{eqnarray}
For this equation a correct initial value problem can be posed ($Q(s=0)=1$). The analysis in a manner similar to that presented above for equation (\ref{E_Q}) gives only one solution $p(s)\approx (3/2) \exp(-3s/2)$, since on our way to this equation we have  eliminated the core (with the differentiation), so the solutions that are valid close to it are not present here. We leave the numerical study of this equation for different values of the coefficients as a task for the future.

The second possibility is to obtain a different version of (\ref{E_Q}) in which the turbulent kinetic energy is presented by itself (\ref{vt2-p(s)}) not by its functional dependence on the accretion kinetic energy (\ref{vt2-va2}). In this case the invariant is:
\begin{eqnarray}
\label{E_vt2_Q}
T_0 Q^{2\beta} +
A_0 \exp(-2s)Q^{-4}(s) + s \nonumber \\
- G_0 \frac{\exp(s)Q^3(s) + \int_{s}^{s_0} \exp(s')Q^3(s') ds'}{Q(s)} \nonumber \\ -\frac{G_1}{Q(s)}=E_0~,
\end{eqnarray}
where $T_0\equiv (u_0^2/c_{\rm s}^2)(l_{\rm c}/{\rm pc})^{2\beta}$ is the ratio of the turbulent kinetic energy per unit mass of the fluid element at the boundary of the cloud to the thermal energy per unit mass. The study of the solutions of this invariant for $0\leq \beta \leq1$ would reveal, in the general case, the role of turbulence for the formation of the density distribution and the balance of energies.

A different research path which leaves the frame of the above presented model would be to derive an equation which governs the evolution of the density distribution with time (for time intervals comparable to the lifetime of the MC), as the cloud accumulates mass due to accretion  \citep{VS_10}. This non-stationary problem is considerably more complex. Another possibility is to consider polytropic equation of state $p_{\rm th}\propto \rho^\Gamma$, where $\Gamma$ can be lower, equal to or higher than one. The rotation of the core could also be taken into account. Its angular momentum probably plays a vital role for the existence of the second PL-tail with slope $q=-1$.

\section{Conclusion}
\label{Sec-Concl}

In the present work we try to obtain the PDF of the mass density of a spherically symmetric model of MCs, which we call an abstract (or average) representative of a class of MCs which have the same PDF, maximum size $l_{\rm c}$ and temperature $T$. This new concept allows us to study the structure of MCs from a general statistical point of view. The abstract representative of the class that we work with is, from a physical point of view, self-gravitating, isothermal, compressible, turbulent fluid. Starting from the equations for the medium and implementing the assumptions of the model, after averaging over the ensemble of microstates which realize the abstract object, we arrive at two basic equations: the equation of conservation of the energy per unit mass of a fluid element during its motion through the scales of the cloud (\ref{dE/ds=0}) and the equation for the scaling of the accretion velocity (\ref{va2-p(s)}). Then, we express the different types of energies per unit mass that participate in the equation in explicit form and eventually obtain a nonlinear integral equation (\ref{E_Q}) for the function $Q(s)$, from which the PDF $p(s)$ can be easily obtained. The solution of this equation is a though problem. In spite of that, the initial analysis conducted here gives hope that it admits solutions of PL-PDF type that have two slopes, the values of which are in good agreement with previous theoretical studies \citep{Shu_77, Girichidis_ea_14}, simulations \citep{KNW_11, Collins_ea_12} and observations \citep{Kainu_ea_09, Schneider_ea_14, Schneider_ea_15}. In the light of the technical difficulties these primary results encourage us that our statistical model is a step in the right direction.

\section*{Acknowledgements}
The authors would like to thank prof.
 S. Yazadjiev, assoc. prof. A. Zhivkov and prof. I. Uzunov for the valuable discussions. The authors also would like to thank the anonymous reviewer for the valuable comments which improved the manuscript.

\bibliographystyle{mnras}

\end{document}